# Did the Big Bang and cosmic inflation really happen? (A tale of alternative cosmological models)

by Marcin Postolak (April 23, 2024)

*"The evolution of the world can be compared to a display of fireworks that has just ended: some few red wisps, ashes and smoke. Standing on a well-chilled cinder, we see the slow fading of the suns, and we try to recall the vanished brilliance of the origin of worlds."*

**Georges Lemaître**

Dear Reader, sit quietly, close your eyes and try to answer the question of what scientific discipline poses the most fundamental questions... For me, one of such questions is the issue of the origin and beginning (if there was one) of the Universe, that is, of everything that exists and all the laws governing the cosmos. The branch of physics that addresses precisely such topics is ***physical cosmology***, the science about the evolution of the Universe as a whole.

Ever since I can actually remember, I have always been fascinated by the Universe, but not by simply observing the night sky, but rather by trying to understand the laws governing its behaviour. For a very long time, however, I never imagined that this passion would develop into something more than watching programmes on TV and reading popular science books. Throughout my time at school, no one was likely to take my interests seriously, but luckily I am a stubborn and persistent person. In high school I started to think seriously about studying astronomy. After graduating from high school, I decided to start studying astronomy. After a year, I changed my major to theoretical physics. Thus began my journey into the mysteries of the Universe. Let me, my dear reader, try to describe a little of what I have learnt so far...

## The Big Bang... or what all the fuss is about

The ***birth of cosmology as a fully-fledged field of physics*** is considered to be on ***25 November 1915***, when ***Albert Einstein presented the equations of the general theory of relativity*** at a lecture at the Prussian Academy of Sciences in Berlin. It is this theory that makes it possible to study the behaviour of the Universe as a whole.

Edwin Hubble discovered a correlation between distance and recessional velocity in his famous scientific article from 1929 [1]. This linear relationship between the escape velocity of galaxies and their distance is known as the Hubble-Lemaître law. However, the first cosmological model that can be considered the protoplast of today's Big Bang theory is Lemaître's model of the primeval atom [2]. It is worth noting that this paper was published in an unknown journal in French and it went unnoticed in the scientific community for a long time.

cost EUROPEAN COOPERATION IN SCIENCE & TECHNOLOGY

Funded by the European Union

**The Big Bang model** is a hypothesis according to which the evolution of the Universe began with a singular (or near singular) state, followed by a phase of expansion that has continued to the present. Indeed, the term **'Big Bang'** is used to denote the initial moment in the evolution of the Universe. It is a backward extrapolation of the present expansion of the Universe. For this reason, contrary to popular belief, _the Big Bang theory does not explain the origin and very beginning of the Universe_.

According to this model, the (simplified) chronology in the evolution of the Universe presents itself as follows (Fig. 2):

1) **Initial cosmological singularity** - a universe of infinite temperature and density. In fact, we can currently try to describe the cosmos in time after the so-called Planck time, $t_{Pl} \approx 5.39 \times 10^{-44} \ s$ (the smallest time interval that makes physical sense) after the Big Bang;

2) **Cosmological inflation** – speculative (currently best in line with observations) early phase of the Universe evolution in which the cosmos undergoes a rapid exponential expansion. An entire next paragraph is devoted to this stage;

3) **Baryogenesis** - a hypothetical process occurring in the early universe which gave rise to the main constituents of matter - nucleons, i.e. protons and neutrons;

4) **Cooling** - temperature and density of the Universe continue to decrease their values. Symmetry-breaking phase transitions are likely to occur, leading to a separation of fundamental interactions from an initial single unified interaction. Due to the asymmetry between matter and antimatter, only matter particles remain as a result of annihilation. Approximately 380,000 years after the Big Bang, a remnant of the recombination period of electrons and protons is emitted - **Cosmic Microwave Background (CMB)**. CMB seems to be a landmark evidence of the Big Bang theory for the origin of the universe (Fig. 1);

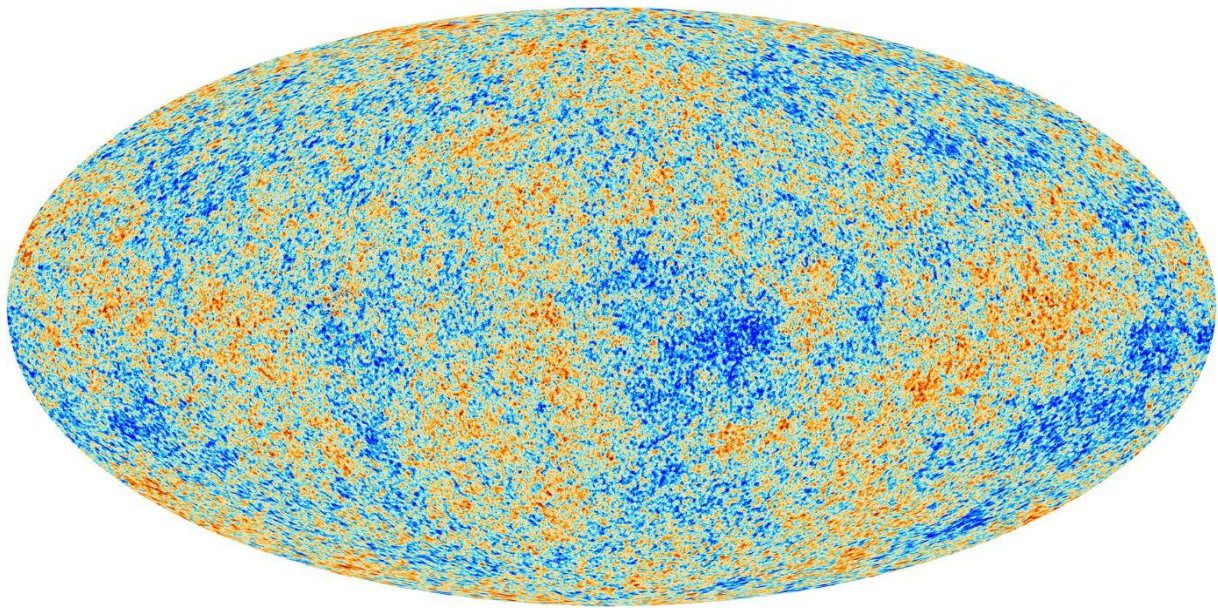

*Fig. 1. CMB map made during the Planck mission (source: ESA/Planck Collaboration).*

5) **Structure formation** - slightly denser areas of evenly distributed matter gravitationally attract nearby matter, with the result that they become even denser. Clouds of gas, stars, galaxies and other astronomical structures observable today are formed. What is important in this

COST EUROPEAN COOPERATION IN SCIENCE & TECHNOLOGY

Funded by the European Union

phase is the process of reionization, which took place between 150 million and 1 billion years after the Big Bang;

6) **Accelerating expansion** - era of the dominance of dark energy (DE) characterised by negative pressure. Assuming that DE is a cosmological constant, the Universe has been undergoing accelerating expansion for about 5 billion years (current estimate of the age of the Universe: 13.79 billion years).

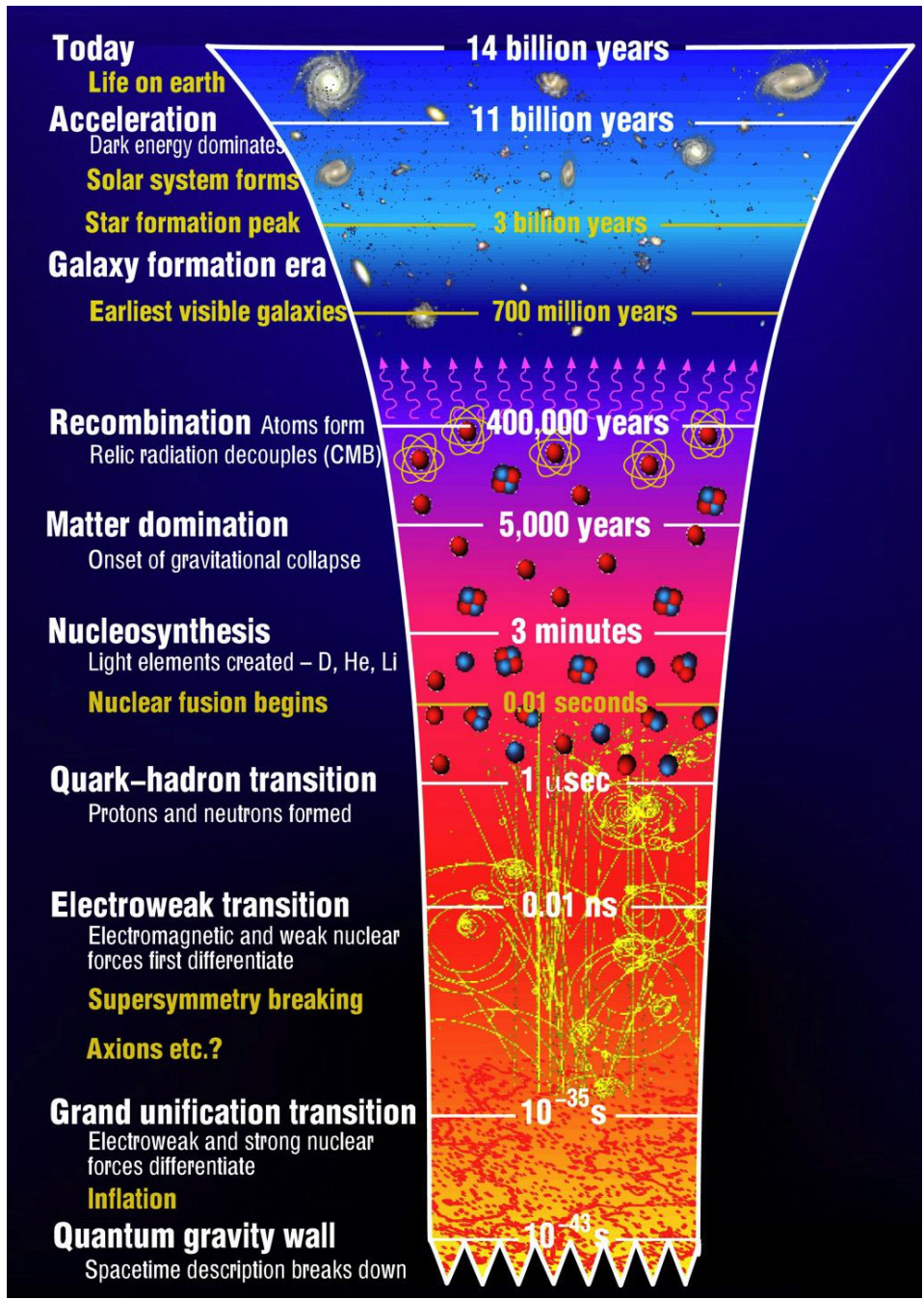

*Fig. 2. Chronology of the Universe according to the Big Bang model (source: Centre for Theoretical Cosmology, University of Cambridge).*

However, the model also has limitations to its applicability and problems associated with it [3,4]. The main ones include [5,6]:

a) ***Initial singularity problem*** - currently known laws of physics break down at the earliest eras of the Universe's existence, and for this reason we do not know whether or not the Universe actually began with an initial singularity. The formulation of the theory of quantum gravity may shed new light on this problem;

b) ***Asymmetry between matter and antimatter*** - we do not know the reason for the breaking of certain types of symmetry in physics, which resulted in the appearance of far more matter than antimatter in the Universe;

c) ***Horizon problem*** - standard physical cosmology cannot explain the homogeneity and isotropy of the Universe at cosmological scales. Without an additional physical mechanism, it is impossible for the CMB map to be so homogeneous (with a factor of $10^{-5}$). Photons at the two extreme 'edges' of the Universe cannot be in a causal relation;

d) ***Flatness problem*** - according to current observational data, the Universe is spatially flat (space can be described by Euclidean geometry), but this is only possible if its density is *almost exactly equal to the so-called critical density* (with an accuracy of at least $10^{-56}$ [7]). Even a minimal change in the value of this parameter would change the spatial geometry (and thus the evolution of the Universe);

e) ***Magnetic monopoles*** - the ***grand unification theories (GUTs)*** (which attempt to unify all fundamental interactions) predict the presence of topological defects in the spatial structure of the Universe, however, no evidence of such objects has been observed to date;

f) ***Dark matter*** - according to the current mass/energy balance of the Universe, more than 26.5% of the Universe's content is so-called dark energy. Under current models, it forms a so-called halo around galaxies, which is responsible for their gravitational stability. It appears to interact only with gravity. If this is a new type of particle then it is not covered by the Standard Model;

g) ***Dark Energy*** - approximately 68.5% of the current content of the Universe is a component characterised by negative pressure, it causes the current accelerating expansion of the Universe. There are many hypotheses about its nature (may have a fixed value or evolve over time), but its nature is even more mysterious than that of dark matter.

## Rapid early expansion…

One of the most important days in the history of cosmology is 15 January 1981. On that day, Alan Guth's paper on the mechanism of ***cosmological inflation*** was published [8]. In the simplest scenario, this model assumes a *rapid exponential increase in the size of the Universe at the earliest stage of evolution* (for most models in the interval $10^{-36} - 10^{-32}$s) (Fig. 3).

In the standard approach, a scalar field (***inflaton***) with self-interaction potential is responsible for the whole mechanism. The most common approach for scalar field inflationary models is the so-called slow-roll approximation, i.e. the assumption that the potential energy of the inflaton is much larger than its kinetic energy. However, the inflationary theory can also be described by the so-called f(R) theory - a modification of the geometrical part of the general relativity (e.g., Starobinsky inflation [9]). Interestingly, Starobinsky published his paper in 1980, a year before Alan Guth, however, he was not aware at the time that it was actually a proposal for an inflationary model of the Universe.

Funded by the European Union

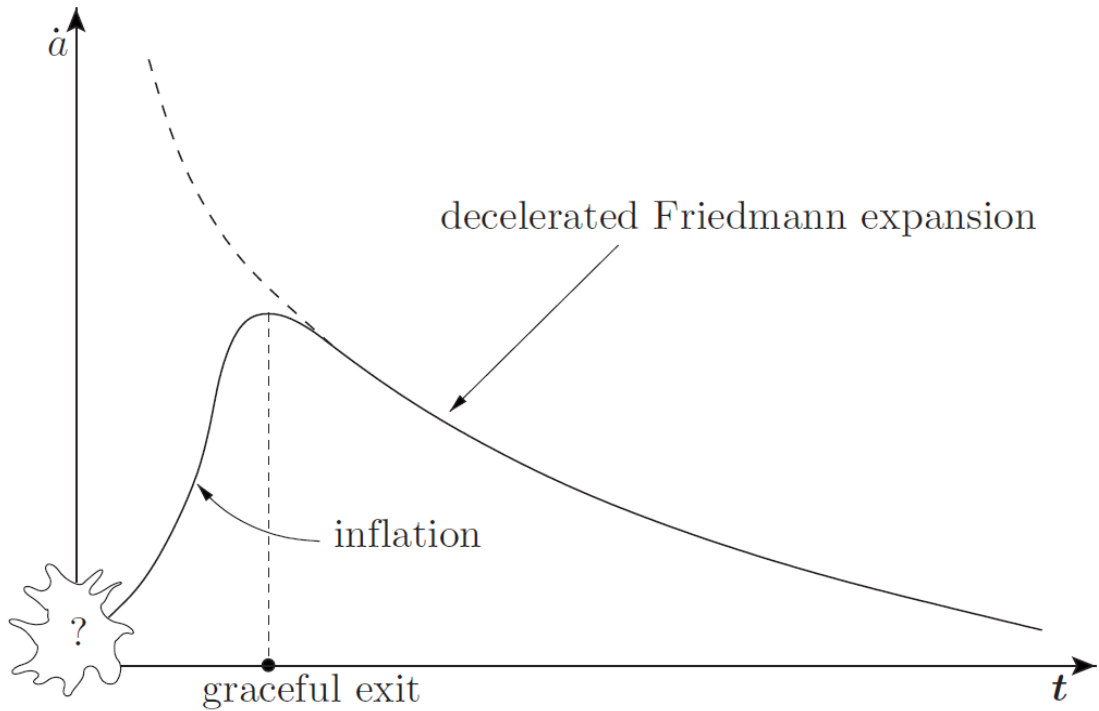

*Fig. 3. The impact of a potential inflationary epoch on the early expansion rate of the Universe* [7].

At the heart of cosmological inflation theory is an attempt to solve the problems plaguing standard Big Bang cosmology. The essence of the explanation of these problems can be presented as follows [7,10,11]:

a) ***Horizon problem solution*** - the observable Universe is homogeneous and isotropic on large scales because it originated from an initially small homogeneous domain that expanded rapidly during the inflationary epoch. Before cosmological inflation, particles could interact with each other due to the fact that *homogeneity scale was always larger than the scale of causality*;

b) ***Flatness problem solution*** - the scalar field (inflaton) energy is approximately constant during the cosmic inflation (unlike matter and radiation). The *exponential increase in the size of the Universe, if it takes long enough, could reduce the near-perfect fit of the Universe's density to the critical density* to today's accuracy of 1%;

c) ***Magnetic monopoles*** - cosmological inflation occurring at a *temperature (energy) lower than that predicted by GUTs* results in all remnants of phase transitions (topological defects, magnetic monopoles, etc.) being very far apart. *Their density is so low that they are currently unobservable*.

At this point, it is worth emphasising that, according to the vast majority of scientists, the standard inflationary model with a single scalar field corresponds very well to observational data (e.g. associated with the CMB [12]). It predicts the formation of the current large-scale structure of the Universe to a good degree. For these reasons, the ***cosmological inflationary model is currently*** *forming a kind of paradigm in physical cosmology* (see, [13]), e.g. it is described in the vast majority of cosmology textbooks.

However, as is the case with most theories and physical models, some scientists have drawn attention to the potential inaccuracies and problems associated with this approach to describing the early Universe, which include [11,14–17]:

- **Nature and origin of the inflaton field** – none of the known scalar fields in theoretical physics correspond to inflationary conditions. Consequently, the following questions can be posed: *What is the nature and origin of the inflationary mechanism?* *Does inflation have an origin related to the new laws of physics?*

- **Graceful entry** – it is a problem associated with the beginning of the inflationary era. The initial conditions of pre-inflationary Universe are claimed to be far from being the standard cosmology (FLRW metric). For this reason, the question arises: *Is inflation a plausible event in a general spacetime (with a general metric)?*

- **Graceful exit and multiverse** - the inflationary model does not directly describe how it terminates, and for this reason, in some varieties of inflationary scenarios, the mechanism appears to have no end or leads to the concept of a multiverse, where this probably leads to the impossibility of observational verifiability (*is this still a science?*). Hence the question can be asked: *What physical mechanism is responsible for an inflation mechanism coming to an end?*

- **Trans-Planckian problem** - analogous to the Big Bang model, potential inflation must have taken place in an interval very close to the Planck time, where the known laws of physics (classical and quantum) do not seem to apply. Therefore, the following question can be formulated: *Can the inflationary model be applied under such initial conditions?*

- **Singularity problem** – in the case of general relativity an initial singularity is inevitable [18], but standard inflation theory does not predict it. For this reason, pre-inflationary period remains unattainable at the level of current cosmological knowledge. Therefore, one may ask: *Did the Universe have its origin in an initial singularity or not?*

- **Fine-tuning and multitude of model variants** - there are a large number of different variants of cosmic inflation scenarios [19] (74 models in this review), so that each version of the model, when confronted with observations, requires the free parameters of the theory to be tuned in different narrow ranges of values. In addition, many scenarios are characterised by a very similar spectrum of perturbations (primordial gravitational waves - PGWs), so even after possible detection of these waves it may not be possible to decide which specific model is correct.

Therefore, in spite of the fact that cosmological inflation is widely accepted, this does not mean that it is a definitively confirmed theory describing the early Universe (at least over a certain time period).

## Why complicate things if we know (not) so much?

Nowadays, physical cosmology is a fully established branch of physics in terms of theory and observations. From a technical point of view, the so-called *Λ-CDM model (LCDM)* is nowadays considered as the 'concordance model' of relativistic cosmology. It is based on the general theory of relativity extended by an additional mechanism – cosmological inflation, which, I hope, is at least partly understandable to you from a conceptual point of view. This model states that the Universe is currently composed of a small amount of radiation, baryonic matter ('ordinary matter' of which we are made) and a dark sector of the Universe (which is a dominant contribution), i.e. dark matter and dark energy (Fig. 4).

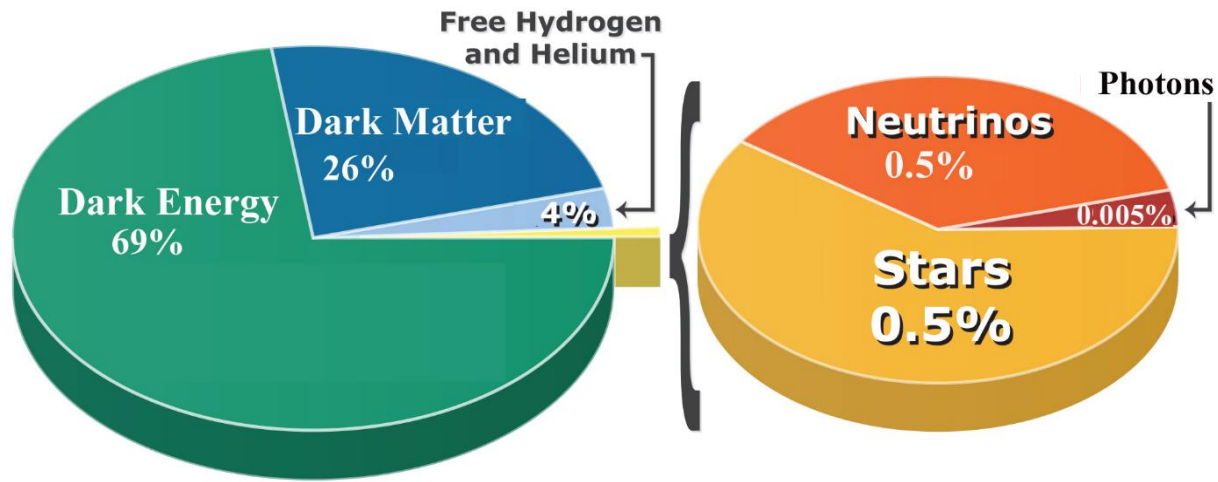

*Fig. 4. Current mass/energy balance of the Universe according to the LCDM model* [73]*.*

In addition to the potential inflation-related problems already mentioned, there is also one that has not been mentioned before. Namely, it is one of the most important unexplained puzzles in fundamental physics so far: cosmological constant problem. In a nutshell, it can be summarised as follows: ***Why is the value of the cosmological constant observable by us (depending on the assumptions made) 60 to 120 orders of magnitude lower than that predicted by quantum field theory*** [20]***?*** The early expansion of the Universe does not seem to answer this vital question. Besides, the attempt to search for new hypotheses can lead to discoveries of new aspects related not only to cosmology, but also to physics in the broadest sense. Ultimately, as long as we do not obtain an unambiguous verification of the correctness of the inflationary scenario (if this is in general possible) then working on other approaches is highly desirable and scientifically necessary.

## …Or is it a collapse after all?

One of the alternatives trying to explain the large-scale structure of the Universe that we observe and its very early evolution are models related to so-called cosmological bounce - ***Bounce Cosmology*** models (Fig. 5). In these, it is assumed that the *Universe undergoes a contraction in the previous stage of evolution followed by a passage through a minimum value of the scale factor (the size of the Universe), either singular or non-singular depending on the model, followed by its expansion and then the evolution we are currently observing*. Therefore, let us give them a little more attention.

Funded by the European Union

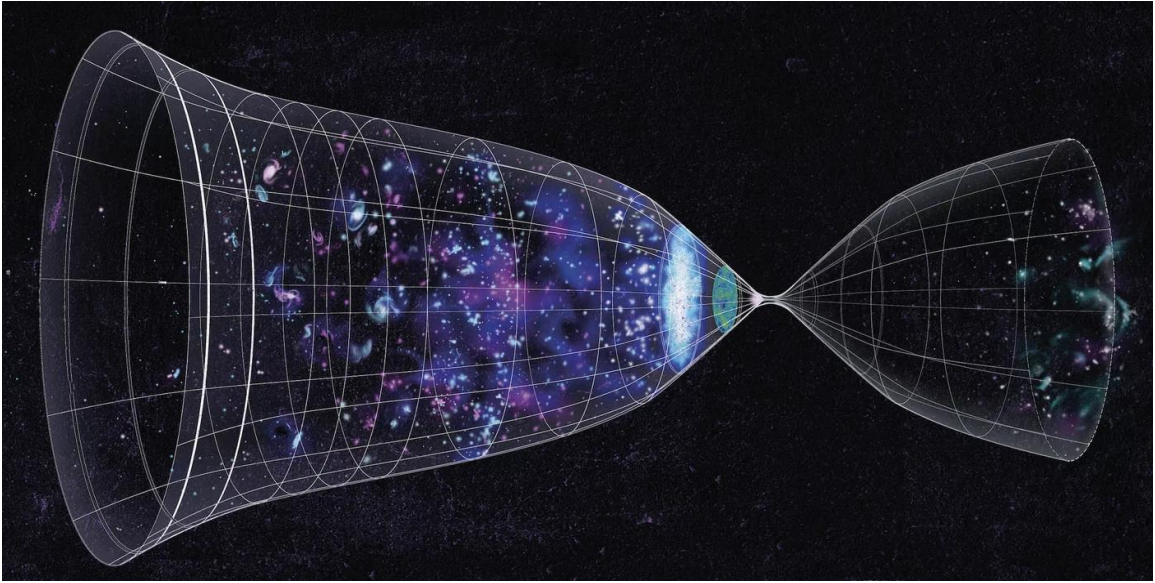

*Fig. 5. Illustration of Universe evolution for cosmological bounce models (source: <u>Max Planck Institute for Gravitational Physics/Anna Ijjas Rosenzweig</u>).*

## A recipe for a (theoretical) description of the bounce phase

One of the main motivations for the introduction of Bounce and Cyclic models is the attempt to remove the initial singularity present in the Big Bang theory formalism. This phenomenon is predicted by **Hawking-Penrose singularity theorems** (see again [18]), which state that the initial singularity is indispensable in the case of an isotropic and homogeneous Universe considered within GR with matter satisfying the so-called **strong energy condition (SEC)**. This condition states that *the sum of the energy density and the pressure produced by the matter must be a non-negative quantity*.

An important aspect is worth noting here. The SEC is one of the four primary **energy conditions** that are intended to be "*reasonable*" (*physically realistic*) constraints on the type of matter/energy present in the Universe under consideration [21]. On the one hand, considering all imaginable cases of matter distribution in the Universe makes Einstein's field equations meaningless, but on the other hand, the fact that these conditions do not follow from any fundamental principle can be considered as a contribution to the discussion on their applicability, at least under very extreme conditions (e.g. just near the Big Bang, at very high densities etc.).

Without going into technical details (which are, however, scientifically very important), physical formalism admits the following formalisms regarding bounce and cyclic models [22–24]:

a) **Modified Matter** - the simplest way is to *consider a scalar field with the opposite sign to the standard one at the component related to its kinetic energy* (e.g. **phantom ghost field** *violating the SEC motivated by the higher-dimensional theories*). In such a case, during a cosmological bounce, the dominant energy contribution of this scalar field is its kinetic energy. This kind of matter can be described by the so-called **scalar-tensor theories (STTs)**, which are a wide class of effective 4D description in cosmology. Furthermore, preliminary studies show that such a formalism *may constitute an attempt to describe the dark matter problem* [25];

b) **Modified Gravity** - modification of the geometrical structure of GR. This modification is most often motivated by the need to incorporate quantum corrections to Einstein's theory relevant in the Planck scale regime. It should be remembered, however, that these theories must reduce to GR in the regime studied observationally by cosmologists;

c) **String Theory** - within this theory, point particles are replaced by strings (1D) or branes (multidimensional) [26]. The starting point is quantum field theory, to which gravitational effects are being tried to be added. The **ekpyrotic scenario** [27,28] is one well-known example of this description (Fig. 6). It is worth noting, however, that despite the mathematical beauty and several decades of research into string theory, *to date there is no observational evidence for its correctness* (it assumes, for example, the correctness of the **supersymmetry (SUSY)** hypothesis as yet unobserved at the LHC);

d) **Quantum Bounces** - a well-known representative of this class of models is **loop quantum gravity (LQG)** or, more precisely, **loop quantum cosmology (LQC)**, whose starting point is the general theory of relativity, to which one tries to add effects related to quantum physics (see, e.g. [29]). LQC states that cosmological singularities could probably be avoided at the quantum level [30].

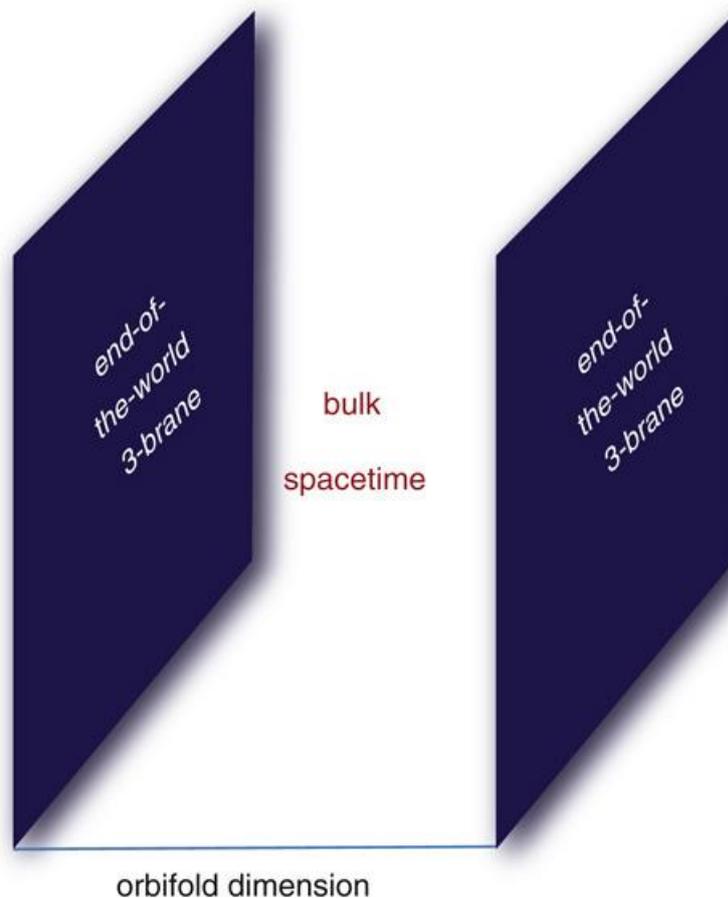

*Fig. 6. The braneworld picture of our universe* [44]. *On one of the branes is located the Universe. The end-of-the-world 3-branes are separated by extra dimensions (bulk spacetime).*

cost
EUROPEAN COOPERATION
IN SCIENCE & TECHNOLOGY

Funded by
the European Union

Before looking a little more closely at some specific examples of the realisation of alternatives to inflation, let us focus for a moment on historical ideas describing cyclic/oscillatory cosmological models.

### Early 20th century historical models

Historically, cyclic and oscillatory models appeared relatively shortly after the publication of Einstein's field equations. Most of them were based solely on certain mathematical assumptions, about which their authors were fully aware that they were not a viable description of our Universe. In this subsection I will present only some of the historical work. A more detailed and complete description of this class of cosmological models can be found in [31].

One of the first approaches to the topic was an article by **Alexander Friedmann**, in which he considered a closed Universe model [32]. However, he only described a single cycle in the evolution of the Universe, i.e. from expansion to contraction. He called the resulting case a *'periodic world'*. Significantly from a physical point of view, he did not consider the laws of thermodynamics at all.

In the 1930s, **Richard Tolman** attempted to apply the laws governing thermodynamics to physical cosmology. On 15 June 1931, a paper treating the problem of formulating the concept of entropy for the Universe as a whole saw the light of day [33]. According to Tolman, the problem lies, in part, in the fact that, according to classical thermodynamics, the **entropy** (in simple terms, *a measure of disorder*) of the Universe should be increasing at an enormous rate, and yet it has not yet reached its maximum value. In a subsequent paper published in the same year, he formulated the *conditions required for the periodic evolution of the Universe* [34]. He concluded that expansion and contraction of the modelled Universe does not cause an increase in entropy and therefore *'could presumably be repeated over and over again'*.

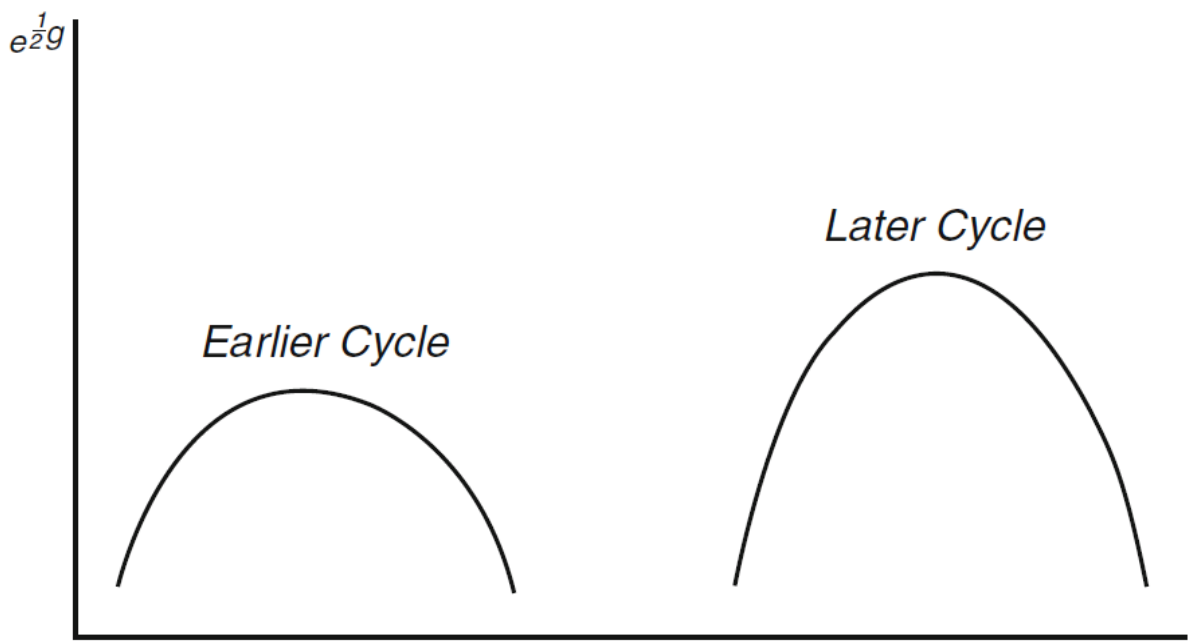

*Fig. 7. Illustration of two cycles of the oscillating Tolman universe, with the later cycle being greater* [36].

He noted, however, that these constraints placed on functions describing the material content of the Universe *do not seem to correspond to any possible physical realisation*. Interestingly, a month before the publication of this study by Tolman, Japanese physicist Tokyo Takeuchi found an example of a solution that met Tolman's conditions [35]. However, this work passed without much notice in the physics community.

Tolman is also the author of *the most important textbook to date on the application of the laws of thermodynamics in terms of relativistic physics and cosmology* [36], in which he states that cyclic evolution is possible if the size of the Universe reaches an increasing maximum size in each successive cycle (Fig. 7).

It is worth mentioning at this point that already *at that time, cyclic cosmological models were subject to quite strong criticism* by scientists in those days, among them **Willem de Sitter** [37], **Howard Robertson** [38] and **Georges Lemaître** [2].

## Examples of specific implementations of bounce and cyclic models

### Matter Bounce

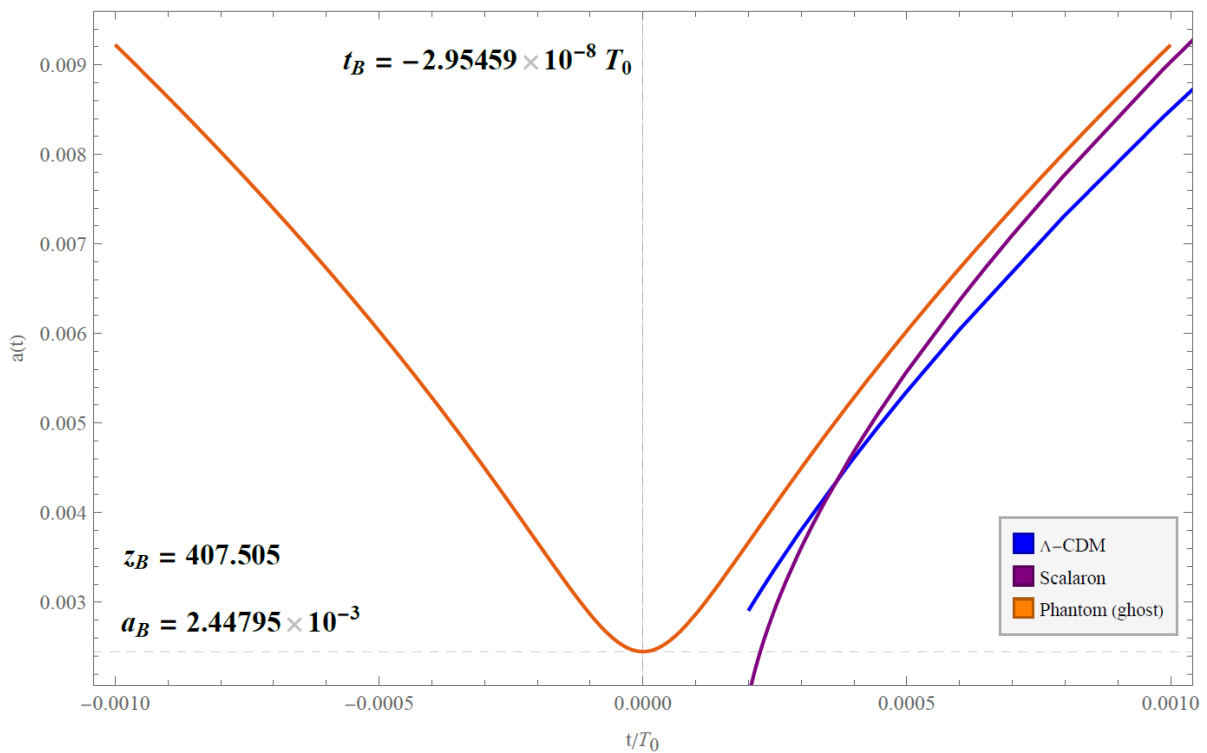

*Fig. 8. Evolution of the scale factor (size of the Universe) in the concrete realization of the matter bounce scenario (with the phantom ghost scalar field)* [25].

In the **matter bounce scenario**, **the exotic form of matter** (e.g., aforementioned kinetic energy of the phantom field in scalar-tensor gravity [25]) **leads to the contraction phase** - the Universe is nearly matter dominated at the very early epoch [39,40]. Then a (non-singular) cosmological bounce occurs, followed by the period of accelerated expansion of the Universe (Fig. 8). Additionally, the model appears to solve the horizon problem.

Furthermore, the phantom (ghost) field in cosmology shows quite interesting properties such as the fact that *its value of the energy density increases with time in the period after cosmological bounce* [25]. It is worth emphasising that the *scalar-tensor cosmological models could in fact be regarded as an effective 4D dimensional description motivated by a more fundamental, so far unknown theory of quantum gravity*.

For this reason, they may represent a meaningful attempt to describe the earliest stages of the evolution of the Universe.

### Ekpyrotic (Cyclic) Universe

The **ekpyrotic Universe** model was originally proposed by **Paul Steinhardt** and **Neil Turok** [41,42] (*for a more recent development, see e.g. the proposal by* **Paul Steinhardt** *and* **Anna (Ijjas) Rosenzweig** [43]). The name of the model is based on the ancient Greek word '*ekpyrosis*' (conflagration), *which refers to Stoic philosophy*.

Instead of inflationary era, ***each cycle consists of a period of slow accelerated expansion followed by contraction that produces the homogeneity, flatness and energy needed to begin the next cycle*** [44,45]. This model can be visualised as the interaction of 2 branes – it is based on M-theory (specific type of string theory). On one of these branes our Universe is located. The approaching and receding of these branes corresponds to the successive stages in the Universe evolution in a given cycle. *Their collision is the moment of "beginning" of the next cycle, or from our point of view, the Big Bang* (Fig. 9).

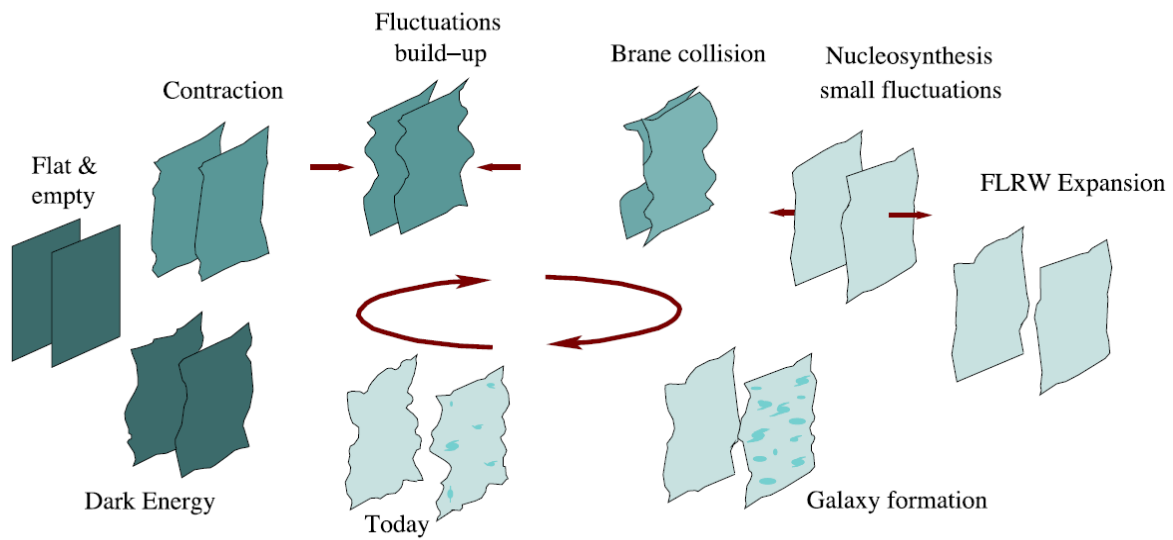

*Fig. 9. Schematic of the cyclic universe scenario* [22]*.*

Each cycle starts with the maximum separation of the branes. Then, as a result of the potential energy, the branes move closer together. As a result of their interaction, density fluctuations are created. In the next step, they collide (not necessarily at the same moment in each fragment) [46]. After this step, standard nucleosynthesis (known from the Big Bang model) occurs. The next stage is a proceeding but non-accelerating expansion. Galaxies form and then the entire large-scale structure of the Universe that we observe. The penultimate epoch is the accelerating expansion of the cosmos (the era of dark energy dominance). The last step in the cycle and at the same time the first step in the next cycle is a flat (spatially) and almost perfectly empty Universe.

In contrast to the inflationary model, in which rapid expansion is responsible for 'smoothing out' the inhomogeneities of the Universe, the *ekpyrotic model in the formalism of scalar-tensor theories uses the negative potential of a scalar field - the negative potential energy of that field (DE equivalent)*. Consequently, *dark energy finds a 'natural' explanation within ekpyrotic theory*. In the vicinity of the bounce, the universe is in the so-called **slow-contraction phase** [47–49], in which the *equation of state parameter (which defines the relation between pressure and density)* takes on values greater than unity - this means that the pressure has significantly larger values than the energy density. Meanwhile, the scalar field itself is equivalent of the natural logarithm from the separation between branes. It is worth pointing out, however, that the use (at least in certain regions) of the negative potential has met with criticism associated with several scientists associated with inflationary theory [50].

This model of a cyclic Universe *differs strongly from the predictions of inflationary cosmology in observational aspects. The primordial gravitational waves (PGWs) that may arise during the ekpyrotic period **are practically impossible to observe***. Their amplitude is much smaller than that of the GWs likely to arise in the early inflationary Universe [51].

## ...Or maybe there was no collapse after all?

### Conformal Cyclic Cosmology

In order to at least partially understand the meaning of *Sir Roger Penrose's* model of **Conformal Cyclic Cosmology (CCC)** [52,53], it is necessary to familiarise oneself with some of the essential foundations of this cosmological scenario. So let us start with the first segment in the model's name, that is, the word '*conformal*'.

cost
EUROPEAN COOPERATION
IN SCIENCE & TECHNOLOGY

Funded by
the European Union

Conformal spacetime geometry

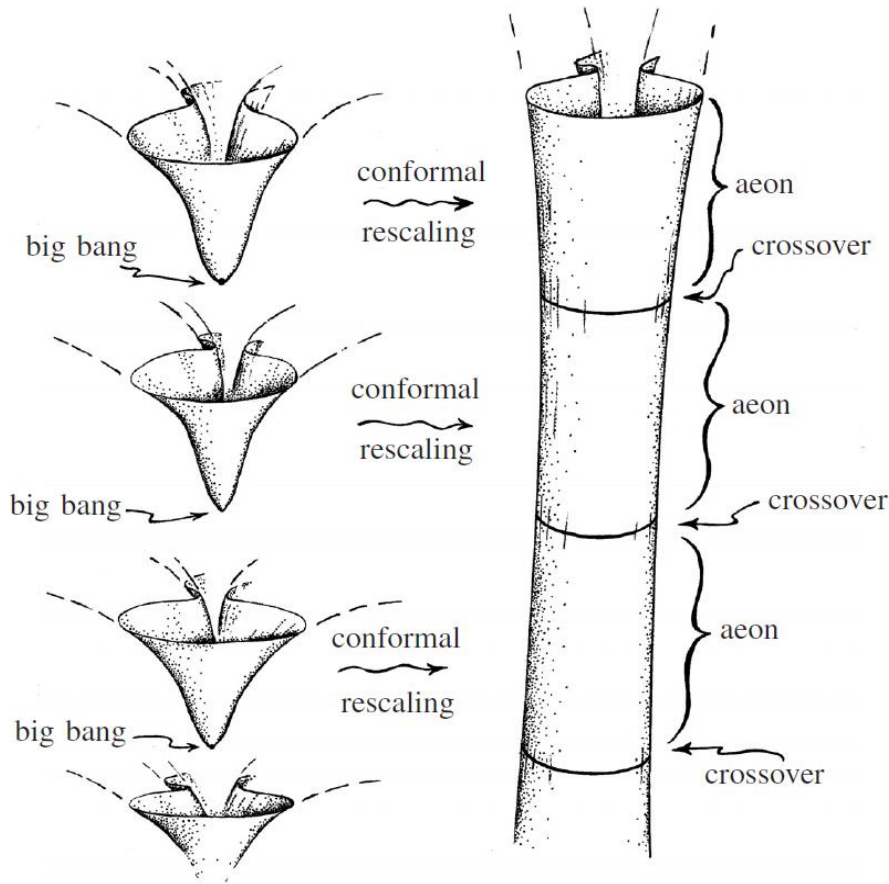

*Fig. 10. Conformal Cyclic Cosmology – aeons visualization [51].*

Let us start with the concept of energy as described in the modern two pillars of physics, i.e. relativistic physics and quantum mechanics. In the ***special theory of relativity (SR)***, *energy is directly proportional to the mass of an object*. In ***quantum physics***, on the other hand, the *energy of a photon is directly proportional to the frequency*. If we compare the two relationships, we conclude that frequency is proportional to mass.

So what can this fact mean? As Roger Penrose expressed it during his Nobel Prize lecture ***'Massive particle is a very perfect clock'***. In such an approach, Universe undergoes repeated cycles of expansion (***aeons***), each starting from its own '*Big Bang*' and finally coming to a stage of accelerated expansion which continues indefinitely (Fig. 10).

It is very important to note that **in this model we do not have any Universe contraction epoch**. If so, the question is: *how is it possible that we are dealing with a cyclic model?* This model is based on the assumption that *each successive aeon of the Universe 'forgets' how big it is*, **both at its Big Bang and in its very remote future it becomes physically identical with the Big Bang of the next aeon**. *This cross-over from aeon to aeon is theoretically possible due to the use of a* **conformal structure**, *that is, in simple terms, one* *that preserves the measures of angles rather than preserving distances*. This conformal structure within Einstein's theory of gravity is a family of **metrics** (*objects describing space-time with a given matter/energy distribution*) that are equivalent to one another via a scale change, which may vary from place to place.

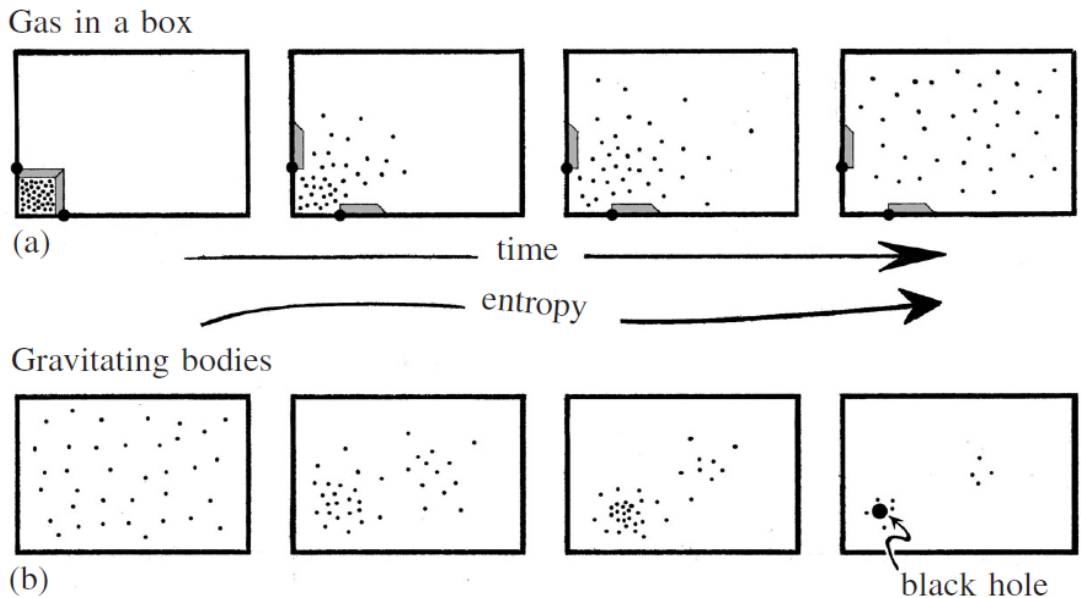

*Fig. 11. Gravity and entropy - In a simple thermodynamic realisation, the gas in a box is spontaneously diluted with time (a). In the case of gravitationally interacting matter (b), on the other hand, we are dealing with the opposite behaviour. Matter forms denser and denser regions with the passage of time, up to the possible formation of a black hole* [52].

### Thermodynamics and the Big Bang

One of the most important laws governing thermodynamics (and, according to many, all of physics) is the so-called **second law of thermodynamics (Second Law)**.

It states that **entropy**, which can also be understood as *an appropriate measure of disorder or lack of "specialness" of the state of the system*, **in closed (isolated) systems must increase with time**.

However, at this point it is necessary to recall the already mentioned problem related to the formulation of the entropy notion for the Universe as a whole and the fundamental question related with the character of the Universe as a physical system, i.e. **is the Universe a closed system?** So far we **do not know the answer** to these questions. Penrose's motivation is the following. According to the Second Law, entropy must decrease in the past time-direction. Therefore, **the initial state of the Universe must be the most special of all**. For this reason, any proposal for the actual nature of this initial state must account for its **extreme specialness** (Fig. 11).

COST
EUROPEAN COOPERATION
IN SCIENCE & TECHNOLOGY

Funded by
the European Union

## The enormity of the specialness and its geometric nature

In order to determine the '*uniqueness*' of the Big Bang, we must use the mathematical definition of entropy. It was introduced by the famous Austrian physicist and philosopher **Ludwig Boltzmann**:

$$S = k_B \ln W.$$

It states that the **entropy (S) of a physical system with a given energy is defined as the quantity proportional to the natural logarithm of the thermodynamic probability (W)**, i.e. the number of microstates available to the system. The constant of proportionality is the Boltzmann constant ($k_B \approx 1.38 \times 10^{-23} \left[\frac{J}{K}\right]$).

In our case, the entire volume might be infinite, as it would be in the case of a spatially infinite Universe. Considering only **baryons**, *the 'ordinary' matter we are made of and see every day*, their amount in the Universe is estimated to be around **$10^{80}$ particles**. The addition of **dark matter** will *increase this value significantly*. The entropy that arises when this number of baryons is collapsed into a BH can be determined by the famous formula determining the entropy of BHs - the **Bekenstein-Hawking formula**, which states that this **entropy is proportional to one quarter of the black hole's event horizon area** (in geometrical units, i.e. for $c = G = 1$). If we consider only baryons then this relationship ultimately yields an estimate of a usable lower bound on the entropy value - **$10^{123}$**.

Based on the above considerations, it is possible to estimate the *ratio of the entire phase space to the Big Bang phase space*. This value must be **greater than $10^{10^{123}}$**. *This is an unimaginably large number*. The conclusion can be only one - **the enormity of the precision in the Big Bang!**

## Problematic assumptions (?)

The cosmological hypothesis of Roger Penrose requires three slightly problematic assumptions:

1) The existence of a **positive cosmological constant $\Lambda$** (*positive value of the vacuum energy density*);
2) After the long duration of the Universe, **all massive particles must disappear or lose their mass (and/or electric charge)**;
3) The validity of the so-called **Weyl Curvature Hypothesis (WCH)** [54].

The first of these does not seem too controversial at first glance, as the concordance model of the Universe (LCDM model) assumes precisely the existence of a positive cosmological constant (dark energy). However, one of the biggest conundrums of science is the question of **whether dark energy is actually non-dynamical like the cosmological constant or whether it is perhaps a time (and space) varying quantity**. *Currently, we do not know the answer to this question*.

The second of these appears to be **the most controversial and problematic**. While the decay of a proton in some extensions of the Standard Model is allowed (but so far this phenomenon has never been observed), **the decay of an electron or its loss of mass and/or electric charge is not proposed in any currently known physical theory**.

As far as the **WCH** is concerned, Penrose argues that the **universe must have initially been in a low entropy state for the Second Law to be fulfilled**. Assuming, as is commonly done, that the matter content of the universe was in thermal equilibrium in the vicinity of the Big Bang, and therefore in a **state of high entropy**, a **low contribution to entropy from the rest of physics** is needed, which means a *contribution from gravity (geometry)*. This implies that the **geometry must be highly organised**. For this mathematical proposal, **we are also not certain that it is correct**.

## Physical implications

There is a very significant debate in the contemporary physics community about the role of internal (mathematical) beauty when it comes to formulating new physical theories [55]. Provided that the role of mathematical symmetries in theoretical physics is unquestionable - **Noether's theorem** (e.g., *the invariance of the laws of physics with respect to time translation gives the energy conservation principle*), the pure mathematical beauty of some physical model proposal does not mean that it has a high probability of being valid at all. **_It is essential to make observational verification of the validity of a given proposal_**.

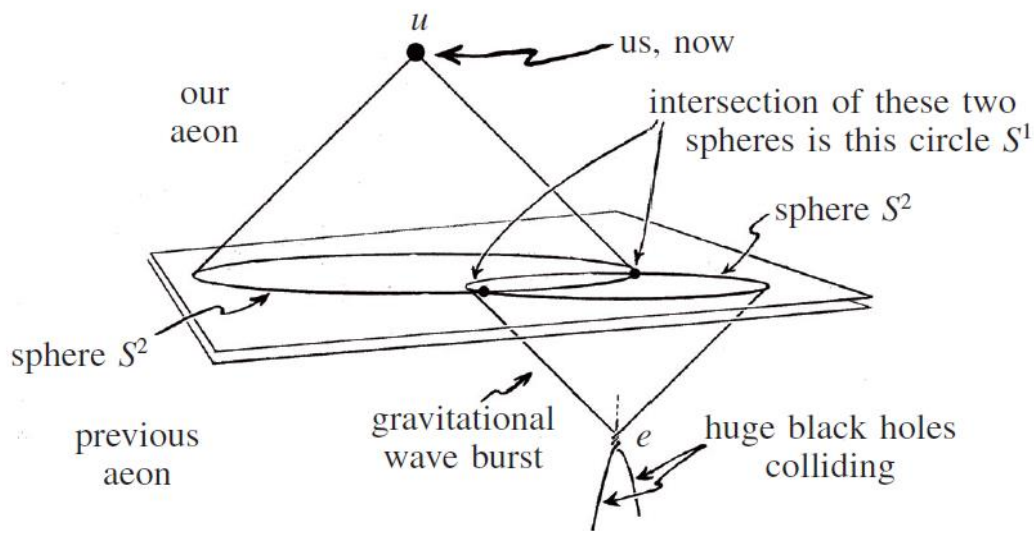

*Fig. 12. Emission of GWs due to collisions of large-sized black holes can cause an imprint in the CMB [53].*

Currently, the concept of so-called "*post-experimental*" science is sometimes formulated in the public domain. According to many scientists, including myself, this approach is completely **_non-scientific_**. In science, there must always be an element of testability of the predictions of a given scientific formalism. In the case of the CCC, these appear to be **footprints in the CMB spectrum from the previous aeon** [56].

Theoretically, they could be formed by the collision of very massive BHs in the previous aeon. Such an event would result in the emission of a huge amount of gravitational radiation - GWs. According to Penrose, these waves could pass from the previous aeon to the next one, which would consequently lead to specific concentric disturbances in the CMB map (Fig. 12). However, the alleged observation of these footprints in the CMB spectrum is **subject to criticism from the scientific community** (e.g. [57]).

## Quantum Gravity…?

The last option for replacing the inflationary paradigm is the attempt to formulate a theory of quantum gravity. As mentioned earlier, there is no fully consistent quantum description of gravity. Due to the fact that I am not an expert in the field of quantum gravity, i.e. I deal on a daily basis rather with modified theories of gravity within the framework of classical physics, in this subsection I will present only two very well-known proposals.

### Hartle-Hawking state (no-boundary wave function)

*James Hartle* and *Stephen Hawking* posed a question about the *conditions that describe the ends of space and time*. In their quite radical proposal, they state that **such ends may not exist at all - time and space may have no boundary with our past** [58]. Consequently, there is no need for further boundary conditions. Hence, the authors of this approach state that **it fully determines the initial state of the Universe** - more precisely, it makes it possible to **determine the wave function** (*a mathematical description of the quantum state of a quantum system*) **of the Universe**.

In essence, the proposal states that the **Universe is completely self-contained**. At first glance, this appears to be a *tautology*, because if such a boundary would exist then it would be necessary to specify the conditions describing it (on both sides of it). In other words, **we would need information from 'outside'** (*whatever that might be*) [58].

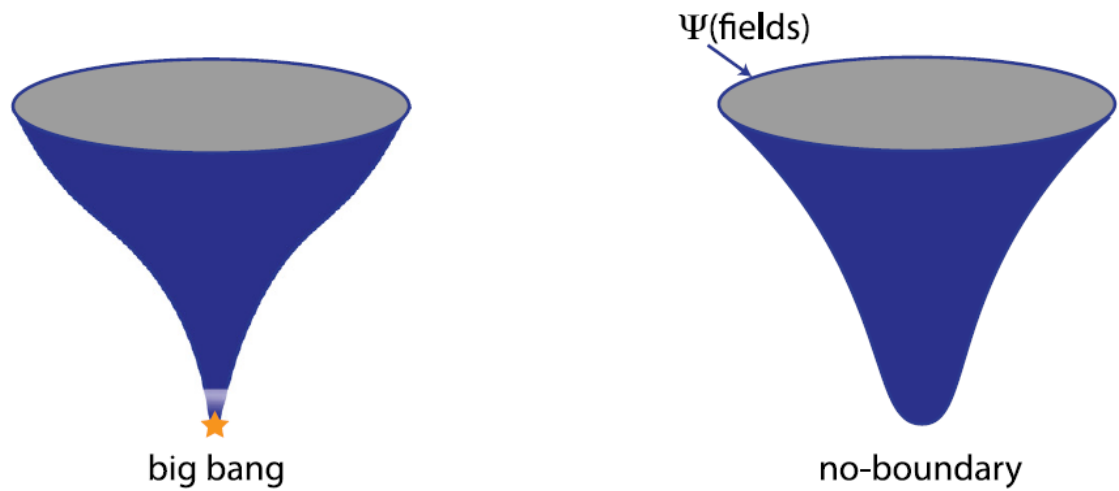

*Fig. 13. Visualization of the no boundary conditions proposal compared to the standard Big Bang scenario* [74]*.*

**How, then, can one actually interpret the Hartle-Hawking proposal?** One interpretation is undoubtedly the absence of an initial singularity, which is replaced by the phenomenon of **quantum tunneling from nothing**, i.e. **from the absence of matter and the lack of spacetime** (Fig. 13).

**This is a fully quantum description of space-time** - when no measurements of it are made (i.e., there are no interactions between it and matter or interactions with itself) **actual spacetime does not exist** [58]. Indeed, due to these interactions, we **perceive spacetime in the classical aspect** described via GR. In contrast, the closer we go back in time to the conventional Big Bang, *the more deviations from classical evolution will occur*.

Funded by the European Union

## Loop Quantum Cosmology

The historical origins of **LQG** theory can be considered to be ***Abhay Ashtekar's groundbreaking work on the implementation of spinorial variables within the GR description*** [59,60]. The main idea behind this approach is an attempt to reconcile Einstein's gravity along with quantum mechanics (quantum field theory) within a single formalism. In other words, the ***formulation of the theory of quantum gravity***. However, unlike the controversial string theory, **LQG is not an attempt to obtain a 'theory of everything'**, i.e., *to unify all interactions within a single theory*.

The starting point for loop quantum gravity is the Einstein's general theory of relativity. To that base, an attempt was made to incorporate quantum physics. However, these theories differ from each other in fundamental aspects. The GR identifies the gravitational field with the space-time metric (a description of the 4D geometry of spacetime) - ***it is a background-independent theory***. In contrast, in the case of **quantum field theory (QFT)**, we are dealing with a **background-dependent description**.

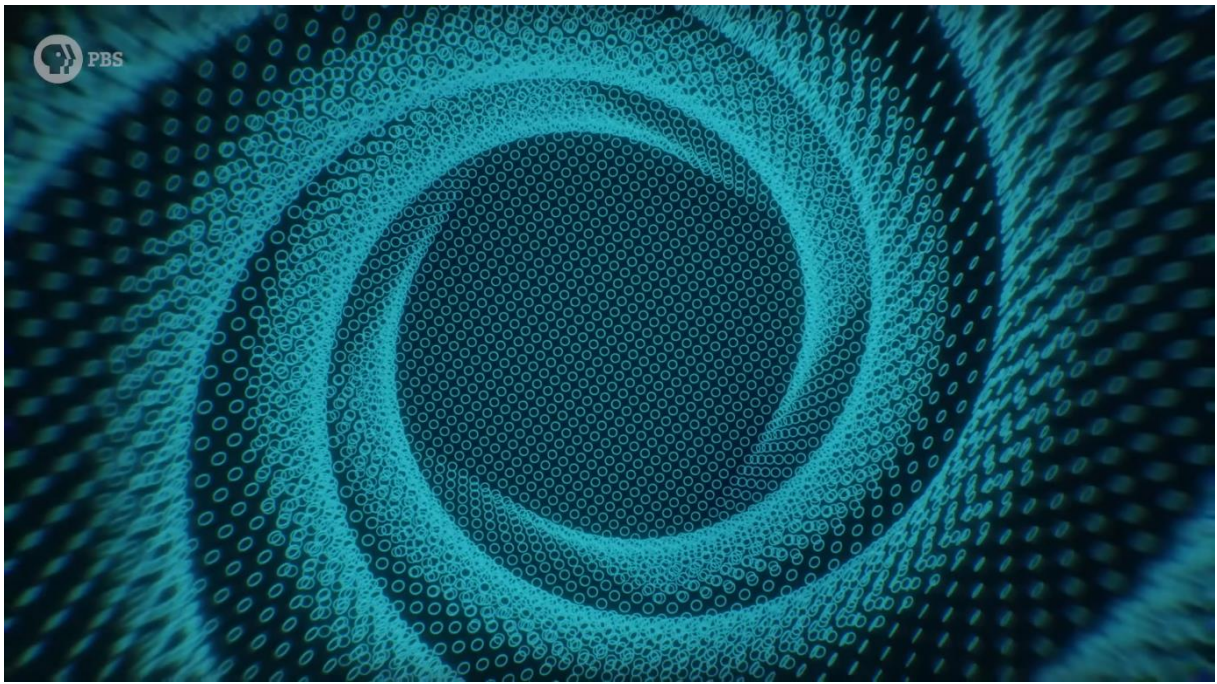

*Fig. 14. Circuits of gravitational field as a representation of the structure of space (source: <u>PBS Space Time YouTube channel</u>).*

***This is the main problem in reconciling the two foundations of modern physics***. This is the reason why LQG was created, it is an ***attempt to quantize the GR in a background-independent level***. The **LQC**, on the other hand, is a ***phenomenological description of cosmology with the inclusion of quantum effects***. It turns out that *quantum effects of geometry* can lead to a *repulsive force*, that is negligible at scales far from the **Planck scale**, but significant at that scale [29]. Furthermore, these sorts of behaviours can result in a ***quantum cosmological bounce*** [30,61,62] (recent developments, see [63,64]). As a result, *LQC also seems to be implementing the Bounce Cosmology scenario*. In contrast to scalar-tensor matter bounce models, there is *no energy conditions violation in this case.*

cost
EUROPEAN COOPERATION
IN SCIENCE & TECHNOLOGY

Funded by
the European Union

At this point, someone could raise a very legitimate question about the name of this theory. Why is it actually called *loop quantum gravity*? **LQG postulates that the spacetime structure is composed of finite loops woven into an extremely fine fabric called spin networks**. Space seems to have an **'atomic' structure** on a distance scale of the order of the **Planck scale** and **is an emergent quantity** (Fig. 14).

As covered in a joint article [65] by **Carlo Rovelli** with **Lee Smolin**: *"If one measured the volume of a physical region or the area of a physical surface with Planck scale accuracy, one would find that any measurement's result falls into the discrete spectra [...]"*.

Thus, we already know what kind of structure space has for this approach to quantizing gravity. Another issue that immediately arises when one tries to quantize gravity is the **problem of time**. In QM time is a variable which is external to the physical system, whereas in GR it is just the another dimension of spacetime . It turns out that in **LQG** and **LQC**, *time is not a fundamental quantity -* **appears as an emergent parameter below some energy scale** [66].

The undoubted successes of loop quantum cosmology include the **emergence of a possible mechanism that produces a cosmological inflation** (the previously mentioned repulsive effects) and the **possible solution to the problem of gravitational singularities** [67].

On the other hand, LQG and LQC can be considered as so-called **'toy models'**, i.e. they are not yet a description of the *full theory of quantum gravity*. This is motivated by the fact that *using truncated classical theory* and then quantizing it can lead to a formalism in which the *complete behaviour of the full theory is not obtained*. Albert Einstein's general theory of relativity seems to be an **effective field theory**, so once it is quantized, it is possible that at the end of the day we will obtain a *description that does not take into account  some fundamental degrees of freedom* [29].

## Doubts and strengths about bounce and cyclic models

Models assuming a non-singular cosmological bounce attempt to address the major challenges facing contemporary physical cosmology. On the other hand, however, they may themselves generate further important problems related to their structure.

### Challenges for Bouncing and Cyclic Cosmologies

The major challenges posed to alternatives hypotheses to cosmological inflation [23]:

- **An attempt to clarify the problems of standard Big Bang cosmology:**
  - **Horizon problem** – all considered models *seem to address this issue*;
  - **Flatness problem** – the matter bounce scenario is *neutral on this issue*, while the ekpyrotic model *explains the problem*;
  - **Entropy** – in the case of the matter bounce and the ekpyrotic Universe *this problem does not occur* - the initial stage is a large and cold Universe;
- **Initial conditions**:
  - The initial conditions of the Universe, without which physicists are unable to determine the exact evolution of the Universe, *should not be restricted to very narrow values*, i.e. **they should not be fine-tuned**;

Funded by the European Union

- o From a technical point of view, it seems that *they should be the same as in the case of inflationary cosmology (the so-called **Bunch-Davies vacuum**)* [68];
- ▪ **Initial inhomogeneities**:
  - o Bounce/Cyclic cosmological scenarios *seem to explain this issue in the opposite way than the cosmic inflation* - instead of expansion, ***slow-contraction** is assumed*;
- ▪ **Anisotropies**:
  - o ***Shear** (components of the energy-momentum tensor - description of the mass/energy distribution in the Universe)* rapidly becomes negligible compared to any other constituent of the Universe, but in general the presence of classical shear leads to uncontrolled growing and this could potentially threatening the whole scenario;
  - o The ***ekpyrotic phase*** may be the solution to this problem;
- ▪ **Relics of the very early Universe**:
  - o In the case of a very young Universe, we are dealing with an ***extremely high energy densities*** ($\rho \sim \rho_{GUT}$ or $\rho \sim \rho_{Planck}$);
  - o High-energy theories predict the formation of ***primordial black holes (PBHs)***, ***topological defects*** and ***exotic types of particles*** at such extremely high energies (e.g. *gravitino in SUSY theory*);
  - o ***Stable relics*** may overclose the Universe and the ***unstable ones*** could interfere with *the cosmic nucleosynthesis*;
  - o The ***problem of magnetic monopoles*** may be avoided in bounce models *only if the maximum temperature of the Universe is below the critical one* (at which symmetry breaking occurs [69]);
- ▪ **NEC and the curvature**:
  - o Many bounce and cyclic models assume that the *Universe is spatially flat in every stage of evolution (including bounce)*, but in the GR formalism this is only possible if the ***null energy condition (NEC)*** *is violated*, i.e. when the sum of energy density and pressure takes on *negative values* - for STTs, this means consideration of *scalar fields with **negative energy***;
  - o At the ***cosmological bounce point***, the famous ***Hubble parameter*** (*which determines the expansion rate of the Universe*) vanishes - the effect of spatial curvature is exactly balanced by the sum of all positive and negative energy contributions;
  - o On the other hand, considering ***non-zero spatial curvature*** can lead to the phenomenon of ***mixmaster Universe*** [70] - *space is subject to contraction and expansion in different directions*, this leads to models of an *anisotropic* and *inhomogeneous* Universe.

## Conclusions

Let us therefore summarise the considerations for models describing the very early Universe, i.e. cosmological inflation and Bounce/Cyclic cosmologies [22–24,71]:

- • ***Inflationary models are self-consistent*** (within the framework of effective theory coupled to GR), which ***cannot (currently) be said of alternative models***;

cost
EUROPEAN COOPERATION
IN SCIENCE & TECHNOLOGY

Funded by
the European Union

- ***Bounce models attempt to explain the problems associated with the initial singularity*** - they go beyond the formalism of the theory of matter coupled to GR (with standard energy conditions);
- ***Cosmic inflation theories may not be consistent from the UV theory point of view*** (general high energy theory must be well-defined at arbitrarily high energies);
- The ***contraction and bounce phase seem to be a natural extension of standard cosmology***, which does not necessarily mean that they describe the real Universe;
- In the Einstein gravity framework, violation of the NEC requires the ***inclusion of unconventional types of matter/energy***, such as ghost scalar field - ***this can result in potential instabilities of the model***;
- ***The theory of the ekpyrotic Universe (at a conceptual level) implies the use of string theory*** (*as yet unproven in any way*) to provide a non-singular behaviour; on the other hand, treating this class of models as STTs may lead to ***treating this approach as an effective description with no connection to higher-dimensional theories***. However, this does not change the fact that ***none of the Bounce and Cyclic Cosmologies models are fully understood***.

## The Universe with or without an initial singularity? That is the question (about the uniqueness of cosmology)

***Physical cosmology*** is a relatively young but very well established discipline of science. Nevertheless, in one very important aspect it differs strongly from other scientific disciplines. Namely, it is about its ***unique nature***.

According to ***George F.R. Ellis***, this uniqueness is related to the following aspects [72]:

1) ***We cannot re-run the universe with the same or altered conditions***;
2) ***We cannot compare the Universe with any similar object***;
3) ***We cannot scientifically establish 'laws of the universe' that might apply to the class of all such objects***;
4) ***Problems arise in applying the idea of probability to cosmology as a whole***;
5) ***We have an essential difficulty in distinguishing between laws of physics and boundary conditions***.

*Does this mean that, from a scientific point of view, we will never know if the Universe had any beginning at all?* **This is very likely to be the case indeed**.

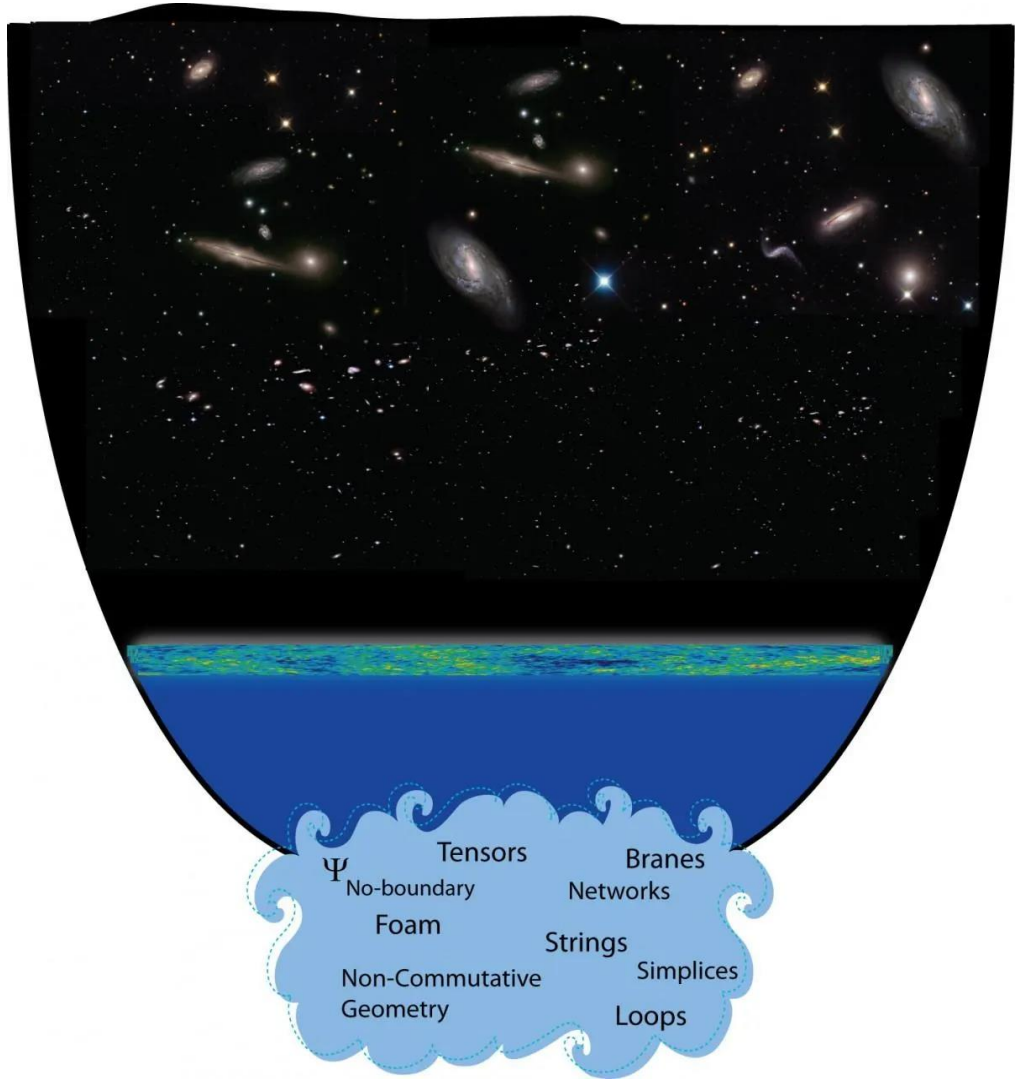

*Fig. 15. Possible solutions to the Universe's initial singularity problem (source: Max Planck Institute for Gravitational Physics, Potsdam).*

However, the goal of science in the broadest sense (including cosmology) is to strive to describe and discover new phenomena and objects. More than once in the history of science there were huge breakthroughs in the understanding of the world (e.g. Darwin's theory of evolution, the discovery of the structure of DNA), phenomena were discovered that no one even supposed to exist (as in the case of Einstein's theory of gravity). Perhaps this will also be the case regarding the description of the earliest stages in the history of the Universe (Fig. 15).

Besides the approaches presented in this text, there are many other attempts to describe the early Universe. Perhaps, any of the ones I have mentioned or even those that do not yet exist will allow us to get closer to one of the greatest puzzles in history - *the question of the nature and origin of the Universe*.

Funded by the European Union

## Take-home message

If someone would ask me what should be kept in mind from the above considerations I would suggest the following crucial facts:

1) Despite several centuries of deliberation on the nature of gravity, to this day ***we do not know its true origin***;
2) The equations of the '*pure*' GR according to ***singularity theorems*** clearly indicate an ***inevitable singularity at the 'birth' of the Universe***;
3) ***Modified theories of gravity*** (e.g., CCC) that point to *significant thermodynamic problems in terms of cosmology* may provide a ***viable alternative*** to the modern standard *Λ-CDM model*;
4) ***Gravitational theories derived from higher-dimensional hypotheses (and their effective 4D descriptions)*** indicate that ***it is possible to avoid the notion of an initial singularity ('beginning' of the Universe)*** and *potentially may be observationally verified in the future*;
5) Current attempts to quantize gravity lead to the ***absence of the initial conditions (Hartle-Hawking state)*** or the ***emergent fuzzy nature of space*** and ***the emergent nature of time (LQG and LQC)***.

Dear Reader, I hope this journey has been at least a little interesting from your point of view. It is obviously not a complete picture of alternative theories to the standard cosmological scenario currently accepted by the physics community. It is a kind of my individual journey through the different areas of cosmological models. I don't know how my scientific 'career' (if it can be so pompously called) will turn out, but of one thing I am sure already. I will always be grateful for the opportunity to learn about the greatest mysteries of modern science - the secrets of the Universe. Science itself, and especially cosmology, teaches humility. It allows one to break away from the difficulties of everyday life into the vastness of the cosmos. Which I wish for everyone as well.

*"Remember to look up at the stars and not down at your feet. Try to make sense of what you see and wonder about what makes the universe exist. Be curious. And however difficult life may seem, there is always something you can do and succeed at. It matters that you don't just give up."*

***Stephen Hawking***

COST
EUROPEAN COOPERATION
IN SCIENCE & TECHNOLOGY

Funded by
the European Union

## Literature

[1]  E. Hubble, *A Relation between Distance and Radial Velocity among Extra-Galactic Nebulae*, Proceedings of the National Academy of Sciences **15**, 168 (1929).

[2]  G. Lemaître, *L'Expansion de l'Espace*, Publications Du Laboratoire d'Astronomie et de Geodesie de l'Universite de Louvain **8**, 101 (1931).

[3]  J. Silk, *The Limits of Cosmology*, in *Why Trust a Theory?: Epistemology of Fundamental Physics*, edited by R. Dardashti, R. Dawid, and K. Thébault (Cambridge University Press, 2019), pp. 227–252.

[4]  C. Smeenk, *Gaining Access to the Early Universe*, in *Why Trust a Theory?: Epistemology of Fundamental Physics*, edited by R. Dardashti, R. Dawid, and K. Thébault (Cambridge University Press, 2019), pp. 315–336.

[5]  R. H. Brandenberger, *Cosmology of the Very Early Universe*, AIP Conf Proc **1268**, 3 (2010).

[6]  P. J. E. Peebles, *Anomalies in Physical Cosmology*, Ann Phys (N Y) **447**, 169159 (2022).

[7]  V. Mukhanov, *Physical Foundations of Cosmology* (Cambridge University Press, 2005).

[8]  A. H. Guth, *Inflationary Universe: A Possible Solution to the Horizon and Flatness Problems*, Physical Review D **23**, 347 (1981).

[9]  A. A. Starobinsky, *A New Type of Isotropic Cosmological Models without Singularity*, Physics Letters B **91**, 99 (1980).

[10]  A. R. Liddle and D. H. Lyth, *Cosmological Inflation and Large-Scale Structure* (Cambridge University Press, 2000).

[11]  R. H. Brandenberger, *Inflationary Cosmology: Progress and Problems*, in *Large Scale Structure Formation*, edited by R. Mansouri and R. Brandenberger (Springer, Dordrecht, 2000), pp. 169–211.

[12]  N. Aghanim et al., *Planck 2018 Results - VI. Cosmological Parameters*, Astron Astrophys **641**, A6 (2020).

[13]  D. Chowdhury, J. Martin, C. Ringeval, and V. Vennin, *Assessing the Scientific Status of Inflation after Planck*, Physical Review D **100**, 083537 (2019).

[14]  R. Penrose, *Difficulties with Inflationary Cosmology*, Ann N Y Acad Sci **571**, 249 (1989).

[15]  A. Ijjas, P. J. Steinhardt, and A. Loeb, *Inflationary Paradigm in Trouble after Planck2013*, Physics Letters B **723**, 261 (2013).

[16]  A. Ijjas, P. J. Steinhardt, and A. Loeb, *Inflationary Schism*, Physics Letters B **736**, 142 (2014).

[17]  M. Jerome, *Cosmic Inflation: Trick or Treat?*, in *Fine-Tuning in the Physical Universe*, edited by D. Sloan, R. Alves Batista, M. T. Hicks, and R. Davies (Cambridge University Press, 2020), pp. 111–173.

[18]  S. W. Hawking and R. Penrose, *The Singularities of Gravitational Collapse and Cosmology*, Proceedings of the Royal Society of London. A. Mathematical and Physical Sciences **314**, 529 (1970).

[19]  J. Martin, C. Ringeval, and V. Vennin, *Encyclopaedia Inflationaris*, Physics of the Dark Universe **5**, 75 (2014).

[20]  S. Weinberg, *The Cosmological Constant Problem*, Rev Mod Phys **61**, 23 (1989).

[21]  E. Curiel, *A Primer on Energy Conditions*, in *Towards a Theory of Spacetime Theories*, edited by D. Lehmkuhl, G. Schiemann, and E. Scholz, Vol. 13 (Springer New York, New York, 2017), pp. 43–104.

[22]  D. Battefeld and P. Peter, *A Critical Review of Classical Bouncing Cosmologies*, Phys Rep **571**, 1 (2015).

[23]  R. Brandenberger and P. Peter, *Bouncing Cosmologies: Progress and Problems*, Found Phys **47**, 797 (2017).

[24]  M. Novello and S. E. P. Bergliaffa, *Bouncing Cosmologies*, Phys Rep **463**, 127 (2008).

[25]  A. Borowiec and M. Postolak, *Is It Possible to Separate Baryonic from Dark Matter within the Λ-CDM Formalism?*, arXiv:2309.10364.

[26]  H. Năstase, *Cosmology and String Theory*, Vol. 197 (Springer International Publishing, Cham, 2019).

[27]  H. Năstase, *The Ekpyrotic Scenario*, in *Cosmology and String Theory*, edited by H. Năstase, Vol. 197 (Springer, 2019), pp. 353–366.

[28]  E. I. Buchbinder, J. Khoury, and B. A. Ovrut, *New Ekpyrotic Cosmology*, Physical Review D **76**, 123503 (2007).

[29]  A. Ashtekar and P. Singh, *Loop Quantum Cosmology: A Status Report*, Class Quantum Gravity **28**, 213001 (2011).

[30]  A. Ashtekar, *Singularity Resolution in Loop Quantum Cosmology: A Brief Overview*, J Phys Conf Ser **189**, 012003 (2009).

[31]  H. Kragh, *Cyclic Models of the Relativistic Universe: The Early History*, in *Beyond Einstein*, edited by D. E. Rowe, T. Sauer, and S. A. Walter (Birkhäuser, New York, NY, 2018), pp. 183–204.

[32]  A. Friedman, *Über Die Krümmung Des Raumes*, Zeitschrift Für Physik **10**, 377 (1922).

[33]  R. C. Tolman, *On the Problem of the Entropy of the Universe as a Whole*, Physical Review **37**, 1639 (1931).

[34]  R. C. Tolman, *On the Theoretical Requirements for a Periodic Behaviour of the Universe*, Physical Review **38**, 1758 (1931).

[35]  T. TAKÉUCHI, *On the Cyclic Universe*, Proceedings of the Physico-Mathematical Society of Japan. 3rd Series **13**, 166 (1931).

[36]  R. C. Tolman, *Relativity, Thermodynamics, and Cosmology*, Vol. 2 (Cambridge University Press (CUP), 1934).

[37]  W. de Sitter, *Some Further Computations Regarding Nonstatic Universes*, Bulletin of the Astronomical Institutes of the Netherlands **6**, 141 (1931).

[38]  H. P. Robertson, *Relativistic Cosmology*, Rev Mod Phys **5**, 62 (1933).

[39]  J. De Haro and J. Amorós, *Viability of the Matter Bounce Scenario*, J Phys Conf Ser **600**, 012024 (2015).

[40]  R. H. Brandenberger, *The Matter Bounce Alternative to Inflationary Cosmology*, (2012).

[41]  J. Khoury, B. A. Ovrut, P. J. Steinhardt, and N. Turok, *Ekpyrotic Universe: Colliding Branes and the Origin of the Hot Big Bang*, Physical Review D **64**, 123522 (2001).

[42]  P. J. Steinhardt and N. Turok, *A Cyclic Model of the Universe*, Science (1979) **296**, 1436 (2002).

[43]  A. Ijjas and P. J. Steinhardt, *A New Kind of Cyclic Universe*, Physics Letters B **795**, 666 (2019).

[44]  J. L. Lehners, *Ekpyrotic and Cyclic Cosmology*, Phys Rep **465**, 223 (2008).

[45]  H. Năstase, *The Cyclic and New Ekpyrotic Scenarios*, in *Cosmology and String Theory*, edited by H. Năstase, Vol. 197 (Springer, 2019), pp. 367–377.

[46]  P. J. Steinhardt and N. Turok, *Endless Universe: Beyond the Big Bang*, 1st ed. (Doubleday, New York, 2007).

[47]  W. G. Cook, I. A. Glushchenko, A. Ijjas, F. Pretorius, and P. J. Steinhardt, *Supersmoothing through Slow Contraction*, Physics Letters B **808**, 135690 (2020).

[48]  A. Ijjas, A. P. Sullivan, F. Pretorius, P. J. Steinhardt, and W. G. Cook, *Ultralocality and Slow Contraction*, Journal of Cosmology and Astroparticle Physics **2021**, 013 (2021).

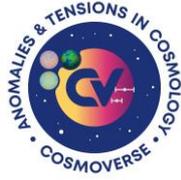

[49] A. Ijjas, W. G. Cook, F. Pretorius, P. J. Steinhardt, and E. Y. Davies, *Robustness of Slow Contraction to Cosmic Initial Conditions*, Journal of Cosmology and Astroparticle Physics **2020**, 030 (2020).

[50] G. Felder, A. Frolov, L. Kofman, and A. Linde, *Cosmology with Negative Potentials*, Physical Review D **66**, 023507 (2002).

[51] L. A. Boyle, P. J. Steinhardt, and N. Turok, *Cosmic Gravitational-Wave Background in a Cyclic Universe*, Physical Review D **69**, 127302 (2004).

[52] R. Penrose, *BEFORE THE BIG BANG: AN OUTRAGEOUS NEW PERSPECTIVE AND ITS IMPLICATIONS FOR PARTICLE PHYSICS*, in *Proceedings of EPAC 2006* (Edinburgh, 2006).

[53] R. Penrose, *Cycles of Time: An Extraordinary New View of the Universe* (The Bodley Head, London, 2010).

[54] R. Penrose, *Singularities and Time-Asymmetry*, in *General Relativity: An Einstein Centenary Survey*, edited by S. Hawking and W. Israel, 1st ed. (Cambridge University Press, Cambridge, 1979), pp. 581–638.

[55] S. Hossenfelder, *Lost in Math: How Beauty Leads Physics Astray* (Basic Books, New York, 2019).

[56] V. G. Gurzadyan and R. Penrose, *On CCC-Predicted Concentric Low-Variance Circles in the CMB Sky*, The European Physical Journal Plus **128**, 1 (2013).

[57] A. Moss, D. Scott, and J. P. Zibin, *No Evidence for Anomalously Low Variance Circles on the Sky*, Journal of Cosmology and Astroparticle Physics **2011**, 033 (2011).

[58] J. B. Hartle and S. W. Hawking, *Wave Function of the Universe*, Physical Review D **28**, 2960 (1983).

[59] A. Ashtekar, *New Variables for Classical and Quantum Gravity*, Phys Rev Lett **57**, 2244 (1986).

[60] A. Ashtekar, *New Hamiltonian Formulation of General Relativity*, Physical Review D **36**, 1587 (1987).

[61] P. Singh, *Are Loop Quantum Cosmos Never Singular?*, Class Quantum Gravity **26**, 125005 (2009).

[62] A. Ashtekar, T. Pawlowski, and P. Singh, *Quantum Nature of the Big Bang*, Phys Rev Lett **96**, 141301 (2006).

[63] M. Kowalczyk and T. Pawłowski, *Regularizations and Quantum Dynamics in Loop Quantum Cosmology*, Physical Review D **108**, 086010 (2023).

[64] M. Bobula, *A Non-Singular Universe out of Hayward Black Hole*, arXiv:2404.12243.

[65] C. Rovelli and L. Smolin, *Discreteness of Area and Volume in Quantum Gravity*, Nucl Phys B **442**, 593 (1995).

[66] S. Brahma, *Emergence of Time in Loop Quantum Gravity*, in *Beyond Spacetime: The Foundations of Quantum Gravity*, edited by N. Huggett, K. Matsubara, and C. Wüthrich (Cambridge University Press, 2020), pp. 53–78.

[67] A. Ashtekar and E. Bianchi, *A Short Review of Loop Quantum Gravity*, Reports on Progress in Physics **84**, 042001 (2021).

[68] T. S. Bunch and P. C. W. Davies, *Quantum Field Theory in de Sitter Space: Renormalization by Point-Splitting*, Proceedings of the Royal Society of London. A. Mathematical and Physical Sciences **360**, 117 (1978).

[69] E. Castellani, *On the Meaning of Symmetry Breaking*, in *Symmetries in Physics*, edited by K. Brading and E. Castellani (Cambridge University Press, 2003), pp. 321–334.

[70] C. W. Misner, *Mixmaster Universe*, Phys Rev Lett **22**, 1071 (1969).

[71] S. Nojiri, S. D. Odintsov, and V. K. Oikonomou, *Modified Gravity Theories on a Nutshell: Inflation, Bounce and Late-Time Evolution*, Phys Rep **692**, 1 (2017).

Funded by
the European Union

[72] G. F. R. Ellis, *The Unique Nature of Cosmology*, in *Revisiting the Foundations of Relativistic Physics*, edited by J. Renn, L. Divarci, P. Schröter, A. Ashtekar, R. S. Cohen, D. Howard, S. Sarkar, and A. Shimony (Springer, Dordrecht, 2003), pp. 193–220.

[73] A. Del Popolo, *The Invisible Universe* (World Scientific, 2021).

[74] J. L. Lehners, *Review of the No-Boundary Wave Function*, Phys Rep **1022**, 1 (2023).